\documentclass[aps,10pt,prd,twocolumn,preprintnumbers,amsmath,amssymb,nofootinbib,superscriptaddress,a4paper,showpacs]{revtex4-1}
\pdfoutput=1
\usepackage[T1]{fontenc}
\usepackage{datetime2}
\usepackage[english]{babel}
\usepackage[utf8]{inputenc}
\usepackage{graphicx}
\usepackage{color}
\usepackage{amsfonts,amsthm}
\usepackage{bm,bbm}
\usepackage{xcolor}
\usepackage{mathrsfs}
\usepackage{amsmath}
\usepackage{float}
\usepackage{pgf,tikz}
\usepackage{mathrsfs}
\usetikzlibrary{arrows}

\def\XM{{\sf X~}}
\usepackage[colorlinks=true, linkcolor=blue, urlcolor=magenta, citecolor=red, anchorcolor=green]{hyperref}

\begin{document}
\title{Positive geometries for all scalar theories from twisted intersection theory}
\author{Nikhil Kalyanapuram}
\email{nkalyanapuram@perimeterinstitute.ca}
\affiliation{Perimeter Institute for Theoretical Physics, 31 Caroline St. N., Waterloo, ON N2L 2Y5, Canada}
\affiliation{Department of Physics and Astronomy, University of Waterloo, Waterloo, ON N2L 3G1, Canada}
\author{Raghav G. Jha}
\email{rjha1@perimeterinstitute.ca}
\affiliation{Perimeter Institute for Theoretical Physics, 31 Caroline St. N., Waterloo, ON N2L 2Y5, Canada}

\begin{abstract}
We show that accordiohedra furnish polytopes which encode amplitudes for all massive scalar field theories with generic interactions. 
This is done by deriving integral formulae for the Feynman diagrams at tree level and integrands at the one-loop level in the planar limit using the twisted intersection theory of convex realizations of the accordiohedron polytopes.
\end{abstract}


\maketitle

\section{Introduction}
\label{sec:intro}
Over the last few years, the study of scattering amplitudes has revealed a number of surprising connections with mathematics. Crucially, deep ties to geometry, topology  and combinatorics \cite{Mizera:2017cqs,Mizera:2017rqa,Mizera:2019gea, ArkaniHamed:2012nw,Arkani-Hamed:2013jha,Arkani-Hamed:2014dca,Arkani-Hamed:2017mur,Arkani-Hamed:2017tmz,Arkani-Hamed:2017vfh,Arkani-Hamed:2019mrd,Arkani-Hamed:2019rds,Banerjee:2018tun,Salvatori:2018fjp,Salvatori:2018aha,Aneesh:2019cvt,Raman:2019utu,Jagadale:2019byr,Witten:2003nn,Roiban:2004yf,Roiban:2004vt,Cachazo:2013gna,Cachazo:2013hca,Cachazo:2013iaa,Cachazo:2013iea,Cachazo:2014nsa,Cachazo:2014xea,Cachazo:2015aol,Mason:2013sva,Adamo:2013tsa,Geyer:2015bja,Geyer:2015jch,Geyer:2016wjx,Geyer:2018xwu,Mastrolia:2018uzb,Frellesvig:2019kgj,Frellesvig:2019uqt} have been established, which have led to the discovery of new ways of computing these quantities. 

In this work, we focus on building upon the seminal developments in the last few years, namely the positive geometry program due to Arkani-Hamed et al. \cite{Arkani-Hamed:2017mur}, and the twisted intersection theory of Mizera \cite{Mizera:2017rqa}. In these works, it was seen that for a wide class of theories built out of trivalent vertices, the planar Feynman diagrams are encoded by the geometry of a polytope known as the associahedron. This was extended to massless scalar theories with generic interactions in \cite{Jagadale:2019byr}, in which a polytope known as the accordiohedron was introduced. In this article, we propose a broad generalization of this line of research by applying the technology of intersection theory to the accordiohedron polytopes.

We seek to address two open questions in the literature. These are as follows. So far, attention has been restricted to the handling of massless interacting particles. The reason for this is the specific realization of the associahedra as convex polytopes, which puts severe restrictions on the masses of the interacting particles. Here, we extend the positive geometry program to all scalar theories while utilizing a convex realization of accordiohedra that removes this restriction on the mass, and are thus able to treat without any difficulty the interactions between particles of arbitrary mass. 

As far as the positive geometry program is concerned, loop effects have been difficult to incorporate. Technical restrictions have forced us to only deal with $\phi^3$ interactions among massless particles at one loop level. We rectify this by proposing a class of accordiohedra which describe interactions between particles in any scalar theory at one loop, in the planar limit. Our construction also allows us to handle different kinds of Feynman diagrams separately, for example, allowing us to treat tadpoles and bubble diagrams distinctly. 

Let us briefly discuss what has been done in the paper and the organization of the text. What has been accomplished is a generalization of the positive geometry framework to take care of massive particles as well. This has been done in section \ref{sec2}. Following this, in section \ref{sec3} we have also described a simple example indicating that the story can be pushed to at least one loop order in arbitrary theories and point out the problems involved in higher loop cases. In doing so, we rectify a problem that has been ignored in the literature, namely the handling of symmetry factors in Feynman diagrams.

\section{Massive $\phi^3+\phi^4$ Scalar Theories}\label{sec2}
In this section, we describe how the twisted intersection theory of accordiohedra can be used to compute scattering amplitudes for generic scalar theories involving massive particles.

Much of the work on positive geometries for scalar theories beyond $\phi^3$ has been done quite recently. For the case of $\phi^4$ and $\phi^p$ interactions, the relevant papers are \cite{Banerjee:2018tun} and \cite{Raman:2019utu} respectively. The formalism for studying generic theories was worked out in \cite{Jagadale:2019byr}. Conspicuously, the analysis in these papers worked specifically for massless particles. 

\begin{figure}[H]
\centering
\includegraphics[width=0.3\textwidth]{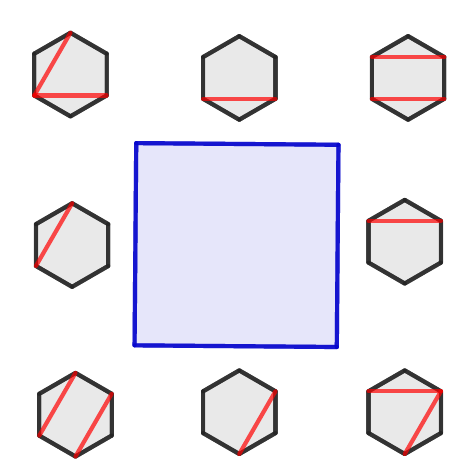}
\caption{Two-dimensional accordiohedron for the reference dissection $(13,46)$. The reference dissection is on the upper right.}\label{fig1}
\end{figure}

In this section, we illustrate how the positive geometry formalism can accomodate massive particles through a development of the intersection theory governing amplitudes in massive scalar theories with $\phi^3 +\phi^4$ interactions. It will turn out that this is the right arena to generalize the study of polytopes controlling these amplitudes for massless particles to massive ones.  To do this, we make use of the accordiohedron data first presented in \cite{Jagadale:2019byr} and the method of realizing these as convex polytopes reviewed in \cite{Kalyanapuram:2019nnf}. To keep the discussion simple, let us restrict ourselves to the case of six particle scattering. This particular process gives rise to two classes of accordiohedra, namely squares and pentagons. Let us begin with the square, which is obtained from the dissection $(13,46)$. The accordiohedron vertices are labelled by $\lbrace{(13,46),(24,46),(26,35),(13,35)\rbrace}$\footnote{We use the notation for dissections set up in \cite{Banerjee:2018tun}. For example, $(13,46)$ means that a hexagon is dissected by drawing a diagonal between vertices $1$ and $3$ and a diagonal between $4$ and $6$.}. Accordingly, the codimension one boundaries are labelled by the partial dissections $\lbrace{(13),(46),(26),(35)\rbrace}$. This is illustrated in Figure (\ref{fig1}).

\begin{figure}[H]
\centering
\includegraphics[width=0.2\textwidth]{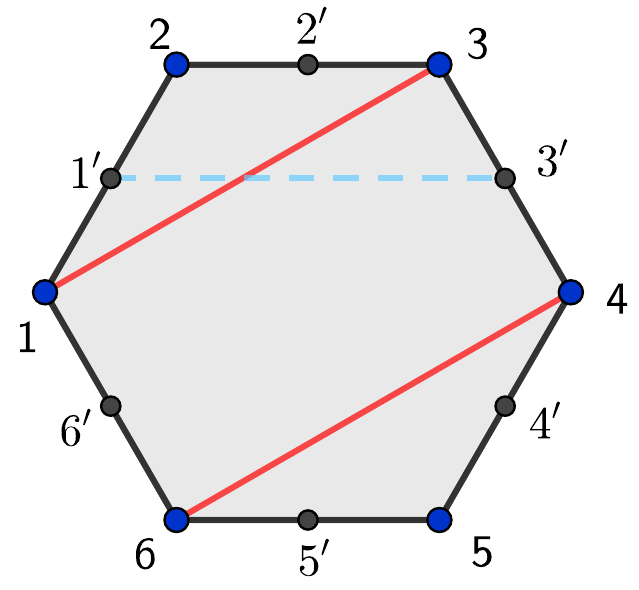}
\caption{The comparison of the dissection $(13)$ (denoted with dashed a line) with the reference $(13,46)$ (denoted with bold red lines).}
\label{fig2}
\end{figure}

The next task is to find a suitable convex embedding of this polytope as a hyperplane arrangement in $\mathbb{CP}^2$, which is rendered possible due to the generic form of the polytopal realization reviewed in \cite{Kalyanapuram:2019nnf}. 
The hyperplanes for an accordiohedron are obtained by comparing the diagonals labelling the facets with the reference dissection. Starting with the facet $(13)$, we have to compare it to the reference dissection (see Figure (\ref{fig2})).

We see from the Figure (\ref{fig2}) that the dissection $(13)$ intersects the reference $(13)$ and forms an inverted ${Z}$ (see Figure 1 of \cite{Thibault:2017nnf}) and does not intersect $(46)$ at all. Using the rules reviewed in \cite{Kalyanapuram:2019nnf}, we can write down the facet $(13)$ (denoted by $f_{1}$) as:

\begin{equation}
    (x\widehat{e}_{13} + y\widehat{e}_{46}) \cdot (\widehat{e}_{13} + 0\widehat{e}_{46}) \leq 1  \implies x \leq 1.
\end{equation}
Here, we have used a basis for $\mathbb{CP}^2$ with basis vectors $\hat{e}_{13}$ and $\hat{e}_{46}$. $x$ and $y$ are the respective values of the inhomogeneous coordinates. Using the same rules, we can now write down the facets $(46)$, $(35)$, and $(26)$ (denoted by $f_{2}, f_{3}, \text{and} ~ f_{4}$) 
as:

\begin{equation}
    \begin{aligned}
    y&\leq 1 \\
    y&\geq -1 \\ 
    x&\geq-1. 
    \end{aligned}
\end{equation}
Clearly, these hyperplanes bound a square. Now, we can shift our interest to the configuration space which is the reference manifold with four hyperplanes above and at infinity removed \emph{i.e.} $\XM = \mathbb{CP}^2 - \bigcup\limits_{i=1}^{4} f_{i} $\footnote{We have not explicitly indicated the hyperplane at infinity, which is formally present. The residue at infinity can be computed by a simple change of variables. It does not however affect our computation of the intersection numbers.}. On this space $\XM$, we define the twist,
\begin{equation}
\begin{aligned}
  \omega_{(13,46)} & =   (X_{13}-m^{2}_{13})d\ln(x-1) + (X_{26}-m^2_{26})d\ln(x+1)\\ &+ (X_{46}-m^{2}_{46})d\ln(y-1) + (X_{35}-m^2_{35})d\ln(y+1)\\
\end{aligned}
\end{equation}
We have used the standard notation to describe generalised Mandelstam variables i.e. $X_{ij}$ is equal to $(p_i + p_{i+1}+...+p_{j-1})^{2}$. These can be visualized as chords of an $n$-gon for an $n$-particle scattering process. Consequently, the dissection $(13,46)$ will translate into the diagram having poles as $X_{13}$ and $X_{46}$ go on-shell. It can be seen from this picture that there are $\frac{n(n-1)}{2}-n$ such variables, which is precisely the dimensionality of the space of Mandelstam variables for an $n$-particle process.

Let us also note the meaning of the notation $m_{ij}^2$. For purposes of maximal generality, we assume that each channel of the scattering process has a different massive pole. $m^2_{ij}$ is the squared mass of the particle propagating along the channel $(ij)$. Now, for the case of a theory with a single kind of particle, all the $m^2_{ij}$ will be equal to $m^2$, where $m^2$ is the mass of the particle. Since we can work out the intersection theory for arbitrary masses, we note that this formalism can be applied for amplitudes such as those in thermal field theories as well, where the $m^2_{ij}$ can be identified with Matsubara frequencies. Thus, we can deal with a fairly wide class of theories using this framework\footnote{We thank an anonymous referee for suggesting that we clarify this point.}.

\begin{figure}[H]
\centering
\includegraphics[width=0.37\textwidth]{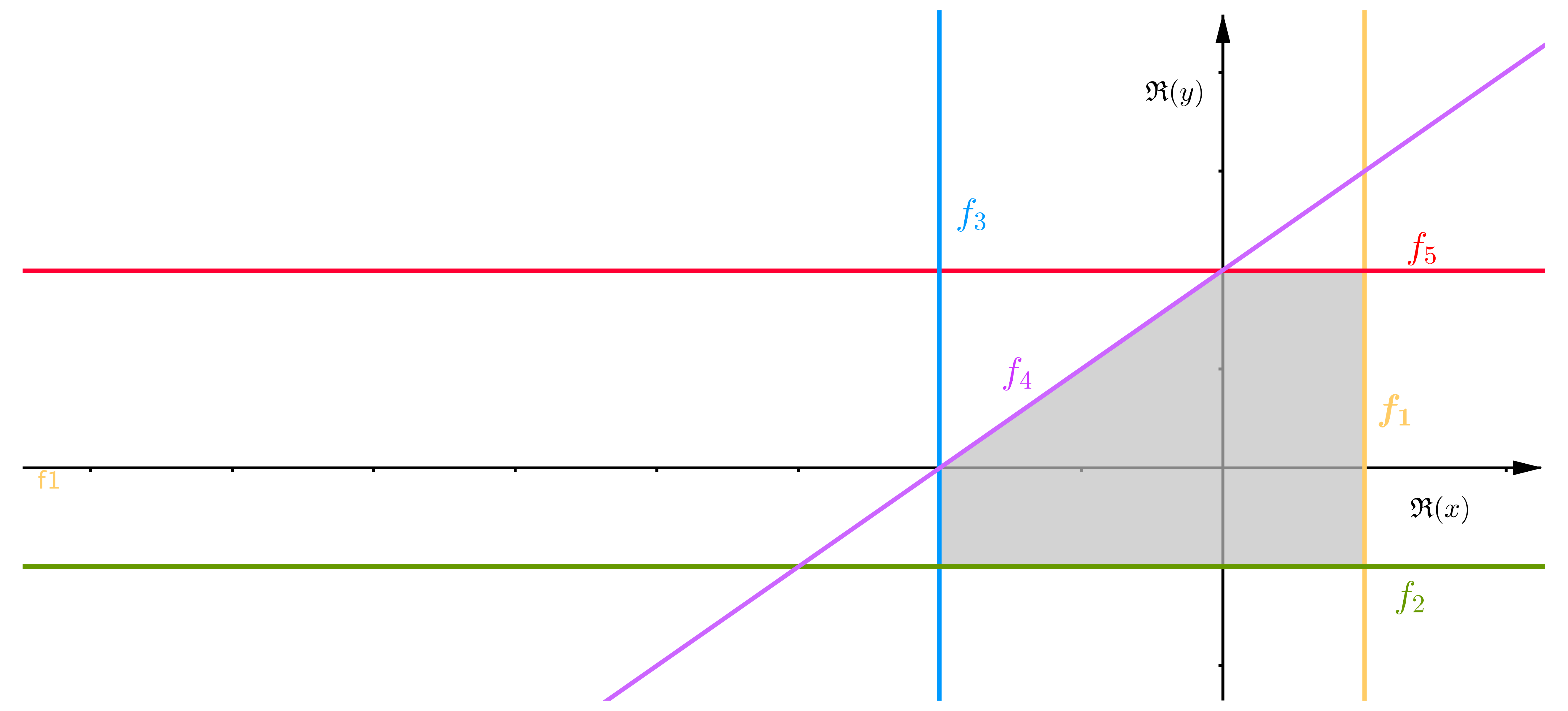}
\caption{Accordiohedron for the dissection $(13,14)$ embedded in $\mathbb{CP}^{2}$.}\label{fig3}
\end{figure}

With this laid out, we can compute the contribution to the scattering amplitude from this polytope by computing the self intersection number of the following form,

\begin{equation}
\begin{aligned}
\varphi_{(13,46)} = &  d\ln f_{1}\wedge d\ln f_{2} + d\ln f_{2}\wedge d\ln f_{3} + \\& d\ln f_{3}\wedge d\ln f_{4} + d\ln f_{4}\wedge d\ln f_{1}.
\end{aligned}
\end{equation}
This can be seen from the formula for intersection numbers, which was first used in the context of scattering amplitudes in \cite{Mizera:2017rqa}. In our case, we are interested in the self intersection number of $\varphi_{(13,46)}$, for which it is sufficient to note that the intersection number is localized on the vertices of the accordiohedron. Schematically, for a given accordiohedron of dimension $n$, if the vertices are labelled by $V_{I}$, the self intersection number of the corresponding form would be given by,

\begin{equation}
    \sum_{f_1\cap...\cap f_{n}=V_{I}}\frac{1}{\alpha_{1}...\alpha_{n}},
\end{equation}
where $\alpha_{i}$ is the weight attached to $f_{i}$. In our case, an application of this formula to $\varphi_{(13,46)}$ gives,

\begin{equation}
\begin{aligned}
&\langle{\varphi_{(13,46)},\varphi_{(13,46)}\rangle}=\\
& \frac{1}{X_{13}-m_{13}^2}  \frac{1}{X_{46}-m_{46}^2} + \frac{1}{X_{46}-m_{46}^2}  \frac{1}{X_{35}-m_{35}^2}\\
 & \frac{1}{X_{35}-m_{35}^2}  \frac{1}{X_{26}-m_{26}^2} + \frac{1}{X_{26}-m_{26}^2}  \frac{1}{X_{13}-m_{13}^2}.
\end{aligned}
\end{equation}
A similar approach can be taken for the pentagon arising from the six particle amplitude in this theory shown in Figure (\ref{fig3}). We will focus on the reference dissection $(13,14)$, which gives rise to a pentagon. The accordiohedron of this reference is labelled by the vertices $\lbrace{(13,14),(24,14),(24,26), (26,36),(13,36)\rbrace}$. The facets may be read off from the set of vertices; they are $\lbrace{(13),(36),(26),(24),(14)\rbrace}$. Using the rules for finding the embedding, we have the following facets (denoted by $f_{1} \cdots f_{5}$.),

\begin{equation}
    \begin{aligned}
x \leq  1 \\
-y \leq  1\\
-x \leq  2 \\
-x+y \leq  2 \\
y \leq  2 .
    \end{aligned}
\end{equation}
These constraints give rise to the 
shaded convex polygon in Figure (\ref{fig3}). 

The kinematical data associated to the amplitude is carried by the twist, which we choose as,

\begin{equation}
    \begin{aligned}
  \omega_{(13,14)}& = (X_{13}-m^{2}_{13})d\ln(x-1) + (X_{14}-m^2_{14})d\ln(y-2)\\ &+ (X_{24}-m^{2}_{24})d\ln(-x+y-2) \\
  &+ (X_{26}-m^2_{26})d\ln(x+2)+ (X_{36}-m^2_{36})d\ln(y+1)\\
\end{aligned}
\end{equation}
Using the assignments for the $f_{i}$'s as defined by the Figure (\ref{fig3}), we compute the self intersection number of the form,

\begin{equation}
\begin{aligned}
\varphi_{(13,14)} &= d\ln f_{1}\wedge d\ln f_{2} + d\ln f_{2}\wedge d\ln f_{3} \\
&+d\ln f_{3}\wedge d\ln f_{4} + d\ln f_{4}\wedge d\ln f_{5}\\
&+d\ln f_{5}\wedge d\ln f_{1},
\end{aligned}
\end{equation}
to compute the amplitude, which becomes,

\begin{equation}
\begin{aligned}
&\langle{\varphi_{(13,14)},\varphi_{(13,14)}\rangle}=\\
& \frac{1}{X_{13}-m_{13}^2}  \frac{1}{X_{36}-m_{36}^2} + \frac{1}{X_{36}-m_{36}^2}  \frac{1}{X_{26}-m_{26}^2}\\
 &+ \frac{1}{X_{26}-m_{26}^2}  \frac{1}{X_{24}-m_{24}^2} + \frac{1}{X_{24}-m_{24}^2}  \frac{1}{X_{14}-m_{14}^2}\\
 &+\frac{1}{X_{14}-m_{14}^2}  \frac{1}{X_{13}-m_{13}^2}
\end{aligned}
\end{equation}

We have indicated that the twist and form are defined for the particular accordiohedron in question by using the subscript $(13,46)$ to denote the reference dissection.

These calculations show that arbitrary mass choices can be made perfectly consistent in the polytopes formalism, even though this aspect is not manifest in the conventional embedding in the kinematical space. It is then obvious that the natural arena for massive scalar theories is twisted intersection theory with a careful convex realization of accordiohedra, which allows us to study scalar theories with arbitrary masses.

We note here that for theories in which a number of massive states can be exchanged in the Mandelstam channels, the amplitude will be given over a sum of intersection numbers; no single intersection number can give all the amplitudes summed over. The two index masses simply provide a general scheme to consider any massive pole structure.

\section{Incorporating Loop Effects} \label{sec3}
A proper discussion including loops while considering scattering amplitudes from the positive geometry viewpoint has been met with some hurdles. For one thing, it has been difficult to make swift progress beyond one loop Feynman diagrams, due to the technical difficulties in dealing with moduli spaces of genus two surfaces. To be precise, these surfaces are not known to be tiled by any regular polytope, making the analysis somewhat tricky. Some progress has been reported at genus one, for which the reader can consult 
\cite{Salvatori:2019phs,Salvatori:2018fjp,Salvatori:2018aha}. 

In addition to the general technical issue of looking at moduli spaces, there is a more mundane issue with including loop interactions. Generically, the integrands appearing in Feynman loop diagrams come with symmetry factors, which encode various degeneracies arising from the large number of ways in which contractions can be performed.

Due to these reasons, it may be more efficient to look at specific classes of Feynman diagrams depending on the nature of renormalization and see if these classes can be described in the polytope framework. To be more concrete, let us consider the case of four-particle scattering (in the planar limit) in $\phi^4$ theory. Here, we receive contributions from two classes of diagrams, namely from diagrams which cause mass renormalization and from  diagrams giving rise to coupling constant renormalization.

\begin{figure}[H]
\centering
\includegraphics[width=0.45\textwidth]{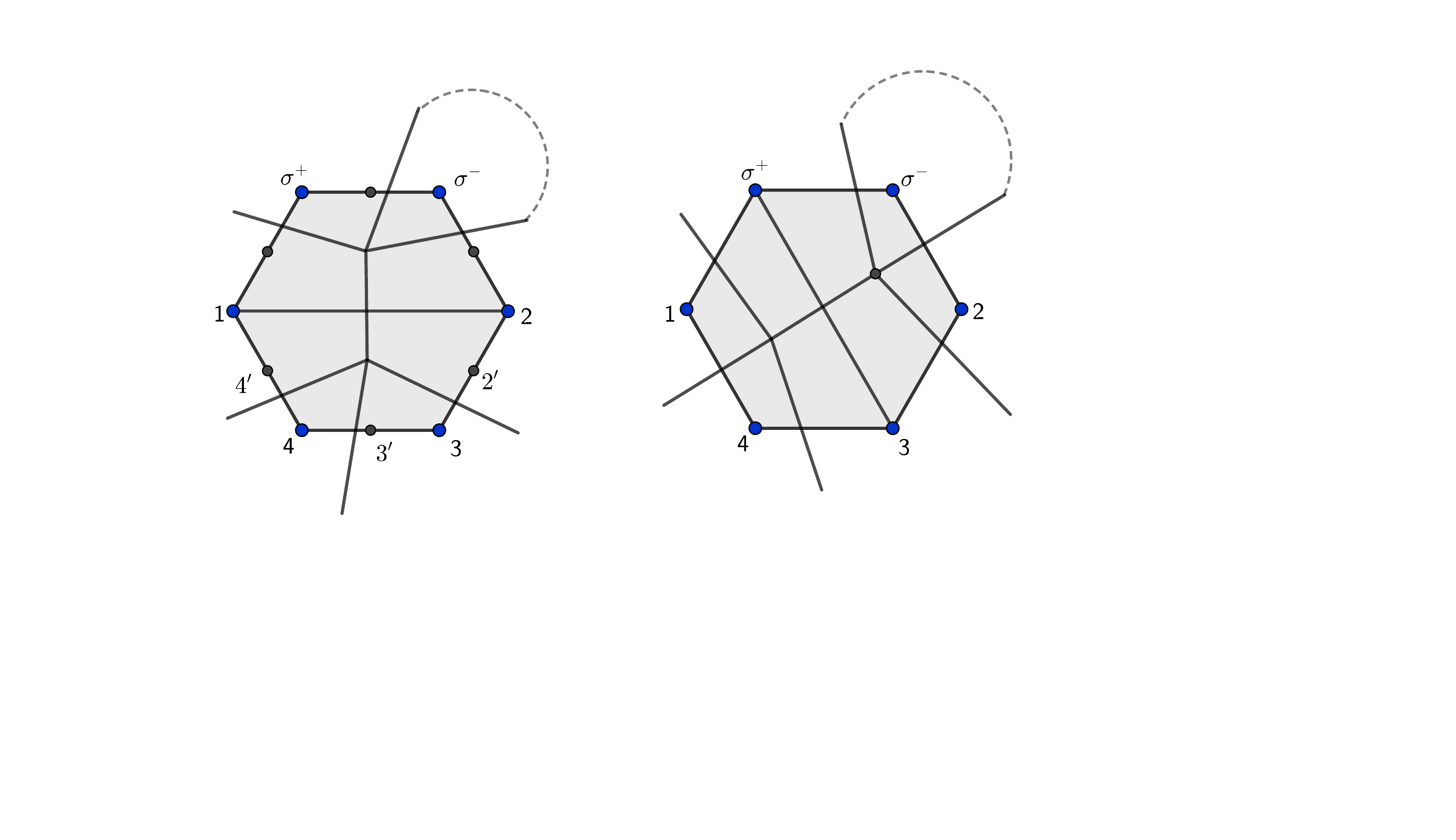}
\caption{The propagator corrections in the $\phi^4$ theory at one loop.}
\label{fig4}
\end{figure}

In order to recast these as intersection numbers, we follow the algorithm that we will now describe. In the field theory limit, which is what we are interested in for the time being, loop interactions are encoded by the complete nodal degeneration of the moduli space $\mathcal{M}_{g,n}$, which is $\mathcal{M}_{0,n+2g}$. Given the $2g$ auxiliary insertions, denoted by $\sigma_{\pm,i}$, with $i$ running from $1$ through $g$, each pair can be sandwiched between a pair of the original insertions as shown in Figure (\ref{fig4}). All possible ways of doing this constitute the tiling of the moduli space. 

Let us specialize to the case of the particle scattering described earlier. Specifically, let the auxiliary points be placed between particles $1$ and $2$. Furthermore, these two insertions are associated with momenta $\ell$ and $-\ell$. If we now look at only the terms giving rise to mass renormalization, we have two diagrams as shown in Figure (\ref{fig4}),

These two diagrams are obtained from the dissections $(12)$ and $(+3)$. Here, $(+3)$ indicates a diagonal between the vertex $\sigma^+$ and $3$. Using these dissections, the technology of accordiohedra and intersection theory may be applied to obtain the stripped integrand, namely the integrand with the loop momentum stripped.

We first find the accordiohedra for the two dissections. For $(12)$, the only compatible dissection is itself. This gives an open accordiohedron, in which the second boundary is pushed to infinity. However, for $(+3)$, the accordiohedron is $\lbrace{(12),(+3)\rbrace}$. Thus, the weights are $0$ and $1$ respectively. We can realize this as $\mathbb{CP}^{1}-\lbrace{0,1,\infty\rbrace}$ with the twist,

\begin{equation}
\begin{aligned}
 \omega_{(+3)} = &(p^2_{1} - m^2)d\ln(x)\\& + (p^2_{2} - \mu^2)d\ln(x-1)
\end{aligned}
\end{equation}

where $m,\mu$ are the masses of virtual particles flowing through the respective channels and the hyperplane at infinity has been indicated. Now, the self intersection number of $\varphi_{(+3)} = d\ln\left(\frac{x}{x-1}\right)$ gives,

\begin{equation}
    \frac{1}{p^2_{1} - m^2} + \frac{1}{p^2_{2} - \mu^2}.
\end{equation}
If the loop momentum is introduced, we get,

\begin{equation}
    \frac{1}{\ell^2}\left( \frac{1}{p^2_{1} - m^2} + \frac{1}{p^2_{2} - \mu^2}\right),
\end{equation}
which is the correct loop integrand. Indeed, this can be absorbed as a renormalization of mass after all channels are taken into account. For this of course, we have to analytically continue past the mass shell, which the intersection theory does not preclude. 

Extending this to loop levels higher than one 
has a technical issue, namely the fact that stripping away all the loop momenta as$\prod_{i}\frac{1}{\ell_i^2}$ is not generically possible, due in large part to the fact that Riemann surfaces of genus $g\geq 2$ can degenerate in very complicated ways to give rise to nodal Riemann spheres\footnote{We are indebted to Sebastian Mizera for clarification regarding this point}. The extension of the results obtained here to higher loop order remains an interesting open problem.

\section{Generic Interactions}
In this section, we briefly describe how the procedure developed above may be applied to generic theories.
Let us first note that the main object of importance is the so called accordiohedron, constructed out of a given set of \emph{dissections}, which label a particular scattering process. Most importantly, these scattering processes can be arbitrarily complicated, so long as the dissections are properly classified and treated appropriately.

Consider for example the rather complicated kinds of polytopes considered in \cite{Raman:2019utu}, in which the accordiohedra for arbitrary $\phi^p$ interactions were obtained. Here, dissections of $p + n(p-2)$-gons into $p$-gons label the collection of all planar Feynman diagrams in an $n$ particle scattering process. Accordingly, the collection of these dissections may be used to obtain the corresponding accordiohedra, which may then be realized as convex polytopes using the methods used here, which were reviewed in \cite{Kalyanapuram:2019nnf}. 

At the same time, we must also bear in mind that there is a practical hurdle to all of this. Leaving aside the computationally intensive aspect, we also remind ourselves that accordiohedra are not generically unique, and a number of distinct accordiohedra usually need to be appropriately weighted and resummed in order to obtain the final amplitude. In our case, this will entail appropriately weighting the corresponding twisted intersection numbers.

From this discussion, the takeaway is simply that the formalism itself can be applied rather straightforwardly, even if cumbersome, such that the real roadblock is to ensure that a self consistent collection of weights can be obtained. Indeed, determining whether or not these weights can be found consistently was an important aspect considered during the work that led to \cite{Jagadale:2019byr}, with some very decent progress also discussed in \cite{Raman:2019utu}. In all the cases considered so far, the weights can be determined consistently. Furthermore, in \cite{Jagadale:2019byr}, it was found that there are at most as many equations determining the weights as there are weights, consequently implying that at least one self consistent solution may be found.

To conclude this section, we remark that the previous points indicate that the procedure outlined in this paper can be carried out for arbitrarily complicated interactions, which although technically challenging at higher points, will always be possible in principle.

\section{Discussion}
In this article, we have developed a framework to handle interactions among scalars in the planar limit which may be arbitrarily complicated from the point of view of twisted intersection theory. Furthermore, we have noted that the formalism presented circumvents some of the rules that are placed on more traditional amplituhedron methods, chiefly among which is the restriction to massless particles. The convex embedding allows for arbitrary choices of mass as well as moving off the mass shell. Among other things, this allows us to treat tadpoles and bubble diagrams with relative ease\footnote{We thank Jacob Bourjaily for discussions on this point.}. Furthermore, we have been able to bring loop amplitudes, at least up to one loop level into the discussion as well while taking care of symmetry factors.

It seems that there are some aspects of this work which can be easily extended. Firstly, in order to keep track of symmetry factors at the loop level, we have by hand restricted to specific subsets of dissections giving rise to loop diagrams according to the nature of renormalization (e.g. mass renormalization and coupling constant renormalization in $\phi^4$ are treated separately.). It remains to be seen whether the symmetry factors and all loop diagrams can be consistently reconciled with one another in the polytopes picture. This seems unlikely, but will surely constitute an interesting future investigation. 

Secondly, it may be interesting to extend our analysis past the realm of scalar theories into richer domains, such as effective field theories (EFT). Historically, the CHY formalism has provided ample insights into EFTs which can be obtained by dimensional reduction of gravity and Yang-Mills. Now, the technology developed here to understand more generic vertices might give us room to look at more exotic EFTs. This is a long-term goal that we hope to pursue in the future.

\section*{ACKNOWLEDGEMENTS} 

We thank Jacob Bourjaily and Sebastian Mizera for going over the draft and helpful comments. We thank Alfredo Guevara and Seyed Faroogh Moosavian for discussions. Research at Perimeter Institute is supported in part by the Government of Canada through the Department of Innovation, Science and Economic Development Canada and by the Province of Ontario through the Ministry of Colleges and Universities.

\bibliographystyle{utphys}
\bibliography{v1.bib}

\end{document}